\documentclass[noshowpacs,pre,aps,superscriptaddress,floatfix,twocolumn]{revtex4}

\usepackage{lineno,hyperref,mathtools,doi}
\usepackage{amsmath}
\usepackage{amssymb,mathrsfs}
\usepackage{graphicx}
\usepackage{hyperref}

\DeclareMathAlphabet{\mathpzc}{OT1}{pzc}{m}{it}
\newcommand{\prt}{\partial}

\newcommand{\la}{\lambda}
\newcommand{\eps}{\varepsilon}

\begin{document}

\title{Wave Breaking in Dispersive Fluid Dynamics
of the Bose-Einstein Condensate\footnote{JETP, {\bf 127,} 903-911 (2018).}}

\author{A. M. Kamchatnov} 
\affiliation{Institute of Spectroscopy,
  Russian Academy of Sciences, Troitsk, Moscow, 108840, Russia}
\affiliation{Moscow Institute of Physics and Technology, Institutsky
  lane 9, Dolgoprudny, Moscow region, 141701, Russia}

\begin{abstract}
The problem of wave breaking during its propagation in the Bose-Einstein condensate
to a stationary medium is considered for the case when the initial profile at the
breaking instant can be approximated by a power function of the form $(-x)^{1/n}$.
The evolution of the wave is described by the Gross-Pitaevskii equation
so that a dispersive shock wave is formed as a result of breaking; this wave can be
represented using the Gurevich-Pitaevskii approach as a modulated periodic solution
to the Gross-Pitaevskii equation, and the evolution of the modulation parameters is
described by the Whitham equations obtained by averaging the conservation laws over
fast oscillations in the wave. The solution to the Whitham modulation equations is
obtained in closed form for $n = 2,3$, and the velocities of the dispersion shock
wave edges for asymptotically long evolution times are determined for arbitrary
integer values $n > 1$. The problem considered here can be applied for
describing the generation of dispersion shock waves observed in experiments with
the Bose-Einstein condensate.
\end{abstract}



\maketitle


{\it Contribution for the JETP special issue in honor of L.~P.~Pitaevskii's 85th birthday}

\section{Introduction}

It is well known that with the disregard of viscosity
and dispersion effects, nonlinear waves experience
``breaking,'' i.e., after a certain critical instant,
the formal solution to corresponding evolution equations
becomes multi-valued (see, for example, \cite{LL-6}).
In classical gas dynamics, this problem is eliminated by taking
into account weak dissipation effects so that instead of
the multi-valuedness domain, a shock wave (i.e., a narrow region
of transition from the flow with some values of parameters
characterizing the flow to a flow
with other values of parameters) appears in the solution.
The width of this transition region is proportional
to coefficients characterizing dissipative processes; in
real conditions, this width is usually of the order of the
mean free path of molecules in the gas. For this reason,
it can be assumed in the macroscopic theory that
this region is a discontinuity in the parameters of the
flow of the medium, and when the medium passes
through the discontinuity, the mass, momentum, and
energy conservation laws must hold. The theory of
shock waves formulated on this basis has been profoundly
developed and has found numerous applications (see, for
example, \cite{LL-6,CF-50}).

In modern physics, however, flows of the medium
in which dissipation processes can be disregarded in
the first approximation are often considered, and then
nonlinear wave breaking is eliminated by taking into
account dispersion effects which lead to formation of
dispersive shock waves (DSWs) (i.e., the evolving
regions of the nonlinear flow of the medium) instead of
the multi-valued domains. Such effects were studied
for the first time in the theory of undular bores in a
shallow water flow (see, e.g., \cite{bl-54}), and the general
nature of this phenomenon was realized by Sagdeev \cite{sagdeev},
who indicated that wave breaking in dispersive
wave systems leads to the formation of an extended
wave structures connecting different states of the flow
like a transition in a shock wave connects different
states of the medium flow with predominant dissipation.
In typical cases, a dispersive shock wave occupies
a spatial region expanding with time so that this wave
is a sequence of solitons at one of its edge and degenerates
into a small-amplitude harmonic wave propagating with the
corresponding group velocity at the
other edge. The main theoretical approach to the
description of DSWs was proposed in the classical
work by Gurevich and Pitaevskii \cite{gp-73} based on the
Whitham theory of modulation of nonlinear waves \cite{whitham}.
In this approach, a DSW is represented in the form of
a modulated periodic solution to the corresponding
nonlinear wave equation, and the slow evolution of the
modulation parameters obeys the Whitham equations
obtained by averaging of the conservation laws over
fast oscillations of physical variables in the wave.
Gurevich and Pitaevskii considered two typical problems
of wave breaking, the evolution of the wave is
described by the Korteweg-de Vries (KdV) equation.
First, a complete analytic solution of the discontinuity
decay was obtained in the case when the initial distribution
at the breaking instant has a sharp jump. Second, they found
the main characteristics of the DSW
in vicinity of the breaking point when the initial distribution
is described by a cubic parabola. Later, Potemin \cite{potemin}
obtained a full analytic solution to this problem
(see also \cite{kamch}). The Gurevich and
Pitaevskii approach to the DSW theory was developed
further and was extended to other equations (see, for
example, review \cite{eh-16}).

One of important applications of the DSW theory
is the dynamics of the Bose-Einstein condensate,
which is described by the Gross-Pitaevskii equation \cite{gross,pit};
for simplicity, we write here this equation in
the standard dimensionless form for a 1D flow of
the condensate:
\begin{equation}\label{eq1}
  i\psi_t+\frac12\psi_{xx}-|\psi|^2\psi=0,
\end{equation}
where $\psi$ is the ``wave function'' of the condensate
flowing along the $x$ coordinate; we assume that the
interaction between atoms is repulsive, which ensures
the stability of its homogeneous state. The theory of
Eq.~(\ref{eq1}) was considered in a huge number of publications.
In particular, its solution in the form of a dark
soliton was obtained in \cite{tsuzuki-1971}, and periodic solutions
were obtained, for example, in \cite{gk-87}. The integrability
of Eq.~(\ref{eq1}) by the inverse scattering transform method
was established in \cite{zs-73} and this approach was used in
\cite{fl-86,pavlov-87} for deriving the modulation equations.
Finally, the problem of the initial discontinuity decay
was analyzed in \cite{gk-87,eggk-95}, and typical wave breaking was
studied in \cite{kku-02}. The apparatus developed in these
works was applied to the description of the DSW
dynamics in the Bose-Einstein condensate.

The DSWs in the condensate were observed experimentally for the first time
in \cite{cornell,hoefer}, where a shock
wave was formed under the action of laser radiation
repelling the condensate. The interpretation of these
observations as DSWs was reported in \cite{kgk-04},
and this experiment was described in \cite{hoefer}
using the theory formulated in \cite{gk-87,eggk-95},
where it was assumed that a DSW is
formed as a result of emergence of a discontinuity.
Although such a discontinuity can be formed when the
flow of the condensate is induced by a piston moving
at a constant velocity \cite{hae-08},
such a case is nevertheless
quite specific, and the wave propagating to the bulk of
a stationary medium in typical situations breaks from
a profile with a certain root singularity rather than a
sharp discontinuity. For instance, a singularity in the
form of a square root appears in the case of the uniformly
accelerated motion of the piston \cite{kk-10}, and it is
clear even from this example that the actual flow of the
condensate may have a quite arbitrary singularity at
the instant of wave breaking. Here, we consider the
DSW formation during wave breaking with an initial
singularity of the type $(-x)^{1/n}$. A detailed theory will be
developed for $n=2$ and $n=3$, and important DSW
characteristics (such as the laws of motion of its edges)
will be obtained for an arbitrary integer $n>1$.  The theory
developed here forms the basis for describing quite
general forms of condensate flow with wave breaking.

\section{Gurevich-Pitaevskii method}

Let us first write the basic relations of the Gurevich-Pitaevskii 
method in the DSW theory as applied
to the dynamics of the Bose-Einstein condensate,
which obeys the Gross-Pitaevskii equation (\ref{eq1}).
It is convenient to represent the periodic solutions to this
equation in terms of more transparent physical variables by
performing the substitution
\begin{equation}\label{eq2}
    \psi(x,t)=\sqrt{\rho(x,t)}\exp\left({i}\int^x u(x',t)dx'\right),
\end{equation}
so that after the separation of the real and imaginary
parts, we obtain the system
\begin{equation}\label{eq3}
\begin{split}
 &\rho_t+(\rho u)_x=0,\\
 &u_t+uu_x+\rho_x+\left[\frac{\rho_x^2}{8\rho^2}
   -\frac{ \rho_{xx}}{4\rho}\right]_x= 0.
\end{split}
\end{equation}
Here, $\rho(x,t)=|\psi(x,t)|^2$  is the condensate density and $u(x,t)$
is the condensate flow velocity. The periodic
solution can be written in the form
\begin{equation}\label{eq4}
\begin{split}
\rho =&\frac14(\la_4-\la_3-\la_2+\la_1)^2+ (\la_4-\la_3)\times\\
&\times(\la_2-\la_1)\,{\rm sn}^2\left(\sqrt{(\la_4-\la_2)(\la_3-\la_1)}\,
\theta,m\right)  ,
\end{split}
\end{equation}
\begin{equation}\label{eq5}
u=V - \frac{j}{\rho}  ,
\end{equation}
where
\begin{equation}\label{eq6}
\begin{split}
j&=\frac{1}{8} (-\lambda_1 - \lambda_2 + \lambda_3 +
\lambda_4)\times \\
&\times(-\lambda_1 + \lambda_2 - \lambda_3 + \lambda_4)
(\lambda_1 - \lambda_2 - \lambda_3 + \lambda_4),\\
\theta&=x-Vt,\qquad V=\frac12 \sum_{i=1}^4\la_i,\\
m&=\frac{(\la_2-\la_1)(\la_4-\la_3)}{(\la_4-\la_2)(\la_3-\la_1)}.
\end{split}
\end{equation}
This solution depends on four parameters $\la_1\leq\la_2\leq\la_3\leq\la_4$,
in terms of which the main characteristics of
the wave can be expressed. In particular, the wavelength is given by
\begin{equation}\label{eq7}
L= \frac{2{K}(m)}{\sqrt{(\la_4-\la_2)(\la_3-\la_1)}},
\end{equation}
where $K(m)$  is the complete elliptic integral of the first
kind. In the limit $\la_3\to\la_2$,  when $m\to1$ and $L\to\infty$,
the periodic wave is transformed into the soliton solution
\begin{equation}\label{eq8}
\rho=\rho_0 - \frac{a_s}{\cosh^2(\sqrt{a_s}(x-V_st))} ,
\end{equation}
where background density $\rho_0$, along which the dark
soliton propagates, its amplitude $a_s$  and velocity $V_s$ are
given by
\begin{equation}\label{eq9}
\begin{split}
\rho_0&=\frac{1}{4}(\la_4 - \la_1)^2,  \quad a_s=(\la_4 - \la_2)(\la_2 - \la_1) ,\\
    V_s&=\frac{1}{2}(\la_1+2\la_2+\la_4) .
\end{split}
\end{equation}
In the opposite limit $\la_3\to\la_4$, when $m\to0$, the wave
amplitude tends to zero, and it is transformed into a
linear harmonic wave propagating over the background with constant density $\rho_0$.

DSW parameters $\la_i$ become slow functions of $x$ and $t$, which change little
in one wavelength $L$. Therefore, we can average the conservation laws for
Eq.~(\ref{eq1}) over fast oscillations in the wave and obtain as
a result the Whitham equations for modulation parameters $\la_i$.
These equations can be written in the form
\begin{equation}\label{eq10}
\frac{\partial \la_i}{\partial
t}+v_i(\la)\frac{\partial \la_i}{\partial x}=0, \qquad i=1,2,3,4,
\end{equation}
where the characteristic velocities are given by
\begin{equation}\label{eq11}
\begin{split}
v_i(\la)&=\left(1-\frac{{L}}{\partial_i{L}}
\partial_i\right)V , \quad i=1,2,3,4 \, ,\\
\partial_i&\equiv\partial/\partial \la_i .
\end{split}
\end{equation}
The substitution of expressions (\ref{eq6}) and (\ref{eq7}) into these
formulas gives the following expressions for velocities:
\begin{equation}\label{eq12}
\begin{split}
v_1&=\frac12 \sum \la_i
-\frac{(\la_4-\la_1)(\la_2-\la_1)K}{(\la_4-\la_1)K-(\la_4-\la_2)E},\\
v_2&=\frac12 \sum \la_i
+\frac{(\la_3-\la_2)(\la_2-\la_1)K}{(\la_3-\la_2)K-(\la_3-\la_1)E},\\
v_3&=\frac12 \sum \la_i
-\frac{(\la_4-\la_3)(\la_3-\la_2)K}{(\la_3-\la_2)K-(\la_4-\la_2)E},\\
v_4&=\frac12 \sum \la_i
+\frac{(\la_4-\la_3)(\la_4-\la_1)K}{(\la_4-\la_1)K-(\la_3-\la_1)E},
\end{split}
\end{equation}
where  $E=E(m)$ is the elliptic integral of the second
kind. Variables $\la_i$ are known as Riemann invariants of
the system of the Whitham modulation equations. We
will also need the limiting expressions for these velocities
at the DSW edges. At the soliton edge, where $\la_3=\la_2$ and $m=1$, we have
\begin{equation} \label{eq13}
\begin{split}
         &v_1= \frac{3}{2} \la_{1} + \frac12{\la_{4}}\, , \quad
         v_4= \frac{3}{2} \la_{4} + \frac12{\la_{1}}, \\
         &v_2=v_3=\frac{1}{2}(\lambda_1+ 2\lambda_2 + \lambda_4),
\end{split}
\end{equation}
while at the small-amplitude edge for $\la_3=\la_4$ and $m=0$ we have
\begin{equation}\label{eq14}
\begin{split}
& v_1= \frac{3}{2} \la_{1} + \frac12{\la_{2}}\, , \quad
v_2= \frac{3}{2} \la_{2} + \frac12{\la_{1}} , \\
   & v_3=v_4=2\lambda_4 +
\frac{(\lambda_2-\lambda_1)^2}{2(\lambda_1+\la_2-2\la_4)}  ,
\end{split}
\end{equation}
(we will not need the expressions for the analogous
limit $\la_1=\la_2$).

In the generalized hodograph method \cite{tsarev} the
solutions to Eqs.~(\ref{eq10})
are sought in the form
\begin{equation}\label{eq15}
    x-v_i(\la)t=w_i(\la),\quad i=1,2,3,4,
\end{equation}
where $v_i(\la)$ are velocities (\ref{eq12}) and $w_i(\la)$ are the sought
functions. If these functions have been determined, $x=x(\la)$ and $t=t(\la)$
turn out to be the functions of parameters $\la_i$. Since these functions
must be inverted and the modulation parameters must
become functions $\la_i=\la_i(x,t)$,
the functions $w_i$ cannot be independent of one another. Differentiating
Eq.~(\ref{eq15}) with respect to $\la_j$, $j\neq i$, and eliminating $t$ from
all pair combinations of the resultant relations, we
arrive at the system of the Tsarev equations
\begin{equation}\label{eq16}
    \frac1{w_i-w_j}\frac{\prt w_i}{\prt \la_j}=\frac1{v_i-v_j}\frac{\prt v_i}{\prt \la_j},
    \quad i\neq j.
\end{equation}
In view of their symmetry in $v_i$ and $w_j$, it is natural to
seek their solution in the form analogous to (\ref{eq11}) (see \cite{gke-92,wright,tian}):
\begin{equation}\label{eq17}
w_i(\la)=\left(1-\frac{L}{\partial_i{L}}
\partial_i\right)W , \quad i=1,2,3,4 .
\end{equation}
Then Eqs.~(\ref{eq16}) are transformed into the system of
Euler-Poisson equations $(i\neq j)$
\begin{equation}\label{eq18}
    \frac{\prt^2W}{\prt\la_i\prt\la_j}-\frac1{2(\la_i-\la_j)}\left(\frac{\prt W}{\prt\la_i}-
    \frac{\prt W}{\prt\la_j}\right)=0.
\end{equation}
For our purposes, it is sufficient to know the set of
solutions obtained from the generating function
\begin{equation}\label{eq19}
    W=\frac{\la^2}{\sqrt{\prod_{i=1}^4(\la-\la_i)}},
\end{equation}
that satisfies Eqs.~(\ref{eq18}) for any $\la$. The expansion of
function (\ref{eq19})
in inverse powers of $\la$  gives
\begin{equation}\label{eq20}
    W=\sum_{k=0}^{\infty}\frac{W^{(k)}}{\la^k}=1+W^{(1)} \frac1{\la}
    +W^{(2)}\frac1{\la^2}+\ldots,
\end{equation}
where $W^{(k)}=W^{(k)}(\la_1,\la_2,\la_3,\la_4)$ are the required
particular solutions to system (\ref{eq18}). As a result, using
expression (\ref{eq17})  we obtain the set of functions $w_i^{(k)}$,
which give the solutions to Whitham equations (\ref{eq10}):
\begin{equation}\label{eq21}
    w_i^{(k)}=W^{(k)}+(2v_i-s_1)\prt_iW^{(k)},
\end{equation}
so that $w_i^{(0)}=1,$ $w_i^{(1)}=v_i$.  The Euler-Poisson equation
is linear in $W$, like expressions (\ref{eq21}) that are linear
in $W^{(k)}$; therefore, any of their linear combinations also
gives the solution
\begin{equation}\label{eq22}
    x-v_i(\la)t=\sum_{k=0}^n A_kw_i^{(k)}(\la),\quad i=1,2,3,4,
\end{equation}
where the number of terms $n$ and constant coefficients $A_k$ are
chosen in accordance with the conditions of the problem.

Let us now show that the above expression of the
DSW theory in the Gurevich and Pitaevskii approach
make it possible to solve the problem of the DSW formation during
wave breaking in the Bose-Einstein condensate.

\section{Dispersionless limit}

Until the instant of breaking, the distributions of
density $\rho(x,t)$ and flow velocity
$u(x,t)$ are smooth functions of spatial coordinate $x$. Moreover, even after
breaking, the DSW occupies a finite spatial region and
its edges at the matching points with the smooth distributions
must be determined as a part of the solution
of the wave breaking problem. In the case of quite
smooth functions $\rho$ and $u$  the terms with a large number
of derivatives in system (\ref{eq3}) can be omitted, which
means the disregard of the dispersion effects; therefore,
the evolution of smooth distributions can be
described by the dispersionless limit equations
\begin{equation}\label{eq23}
    \rho_t+(\rho u)_x=0,\quad u_t+uu_x+\rho_x=0.
\end{equation}
These equations coincide with the ``shallow water''
equations (see \cite{LL-6}), equivalent to the gas dynamic
equations with adiabatic exponent $\gamma=2$. Therefore, their
solutions can be obtained using well-known classical methods.

Equations (\ref{eq23}) can be transformed to diagonal
form by introducing the Riemann invariants
\begin{equation}\label{eq24}
    \la_{\pm}=\frac{u}2\pm\sqrt{\rho},
\end{equation}
so that
\begin{equation}\label{eq25}
\begin{split}
    & \frac{\prt\la_+}{\prt t}+v_+(\la_+,\la_-)\frac{\prt\la_+}{\prt x}=0,\\
    & \frac{\prt\la_-}{\prt t}+v_-(\la_+,\la_-)\frac{\prt\la_-}{\prt x}=0,
    \end{split}
\end{equation}
where
\begin{equation}\label{eq26}
    v_+=\frac32\la_++\frac12\la_-,\quad v_-=\frac12\la_++\frac32\la_-,
\end{equation}
here, $\rho$ and $u$ can be expressed in terms of $\la_{\pm}$ by the
formulas
\begin{equation}\label{eq27}
    \rho=\frac14(\la_+-\la_-)^2,\quad u=\la_++\la_-.
\end{equation}

Generally, both Riemann invariants $\la_{\pm}$ are functions of $x$ and $t$.
We are interested, however, in the
problem in which a wave propagates to the bulk of the
condensate at rest with constant density $\rho_0$. It is known \cite{LL-6},
that only a flow in the form of a simple wave, in
which one of the Riemann invariants is constant, can
border on such a state of the gas. Assuming for definiteness that the wave
propagates to the right, we can
conclude that Riemann invariant $\la_-$ must be constant
and, hence, must have the same value as in the stationary medium bordering the wave:
\begin{equation}\label{eq28}
  \la_-=-\sqrt{\rho_0}.
\end{equation}
Then the second equation in (\ref{eq25}) is satisfied automat-
ically, while the first equation is transformed into the
Hopf equation
\begin{equation}\label{eq29}
  \frac{\prt\la_+}{\prt t}+\left(\frac32\la_+-\frac12\sqrt{\rho_0}\right)
  \frac{\prt\la_+}{\prt x}=0
\end{equation}
with the well-known general solution
$$
x-\left(\frac32\la_+-\frac12\sqrt{\rho_0}\right)t=w(\la_+).
$$
If function $w(\la_+)$ is known, this solution is defined by
the dependence $\la_+=\la_+(x,t)$,
and this function must
be joined at the boundary with the stationary condensate with value
$\la_+=\sqrt{\rho_0}$ in this spatial region.

\begin{figure}[ht]
\centerline{\includegraphics[width=8cm]{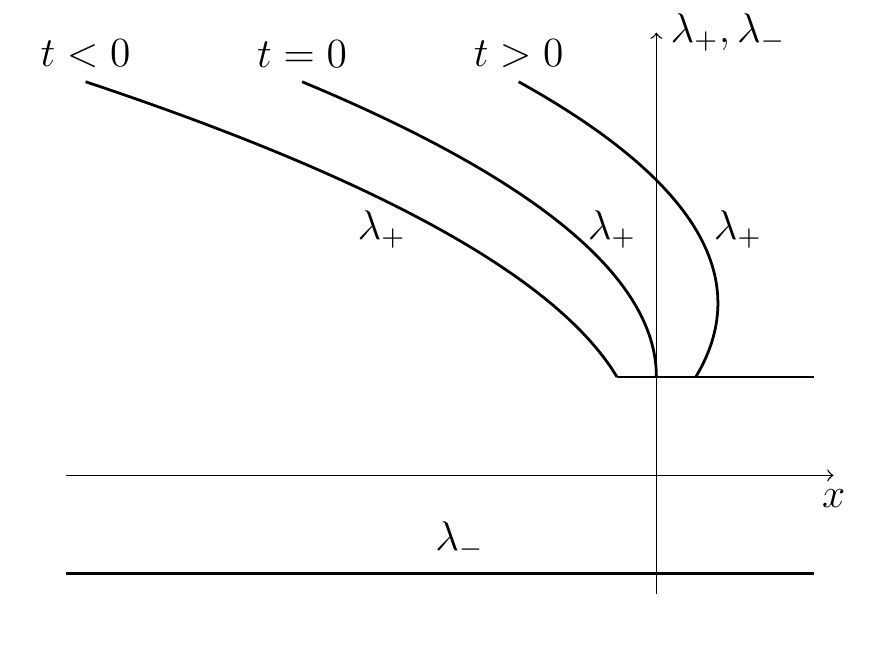}}
\caption{Riemann invariants $\la_{\pm}$
as functions of coordinate $x$ at fixed instant $t$. }
\label{fig1}
\end{figure}

We are interested in the situation when the smooth
solution for $\la_+$ at the breaking instant tends to its
boundary value $\sqrt{\rho_0}$ as a root function of $x$. Choosing
the coordinate system and its origin so that breaking of
Riemann invariant $\la_+$ occurs at instant $t=0$ at the origin $x=0$,
we obtain the dependence
\begin{equation}\label{eq30}
  x-\left(\frac32\la_+-\frac12\sqrt{\rho_0}\right)t=-(\la_+-\sqrt{\rho_0})^n,
\end{equation}
where, to simplify calculations, the units of measurements of length and
time are chosen so that the coefficient on the right-hand side be equal to unity.
Therefore, the dependence of $\la_+$ on $x$ for $t<0$
has no singularities, while, at $t=0$ the root singularity appears,
\begin{equation}\label{eq31}
\la_+=\sqrt{\rho_0}+(-x)^{1/n},
\end{equation}
and the dependence of $\la_+$ on $x$ for $t>0$
becomes multi-valued (see Fig.~\ref{fig1}).

\section{Dispersive shock wave}

\begin{figure}[ht]
\centerline{\includegraphics[width=8cm]{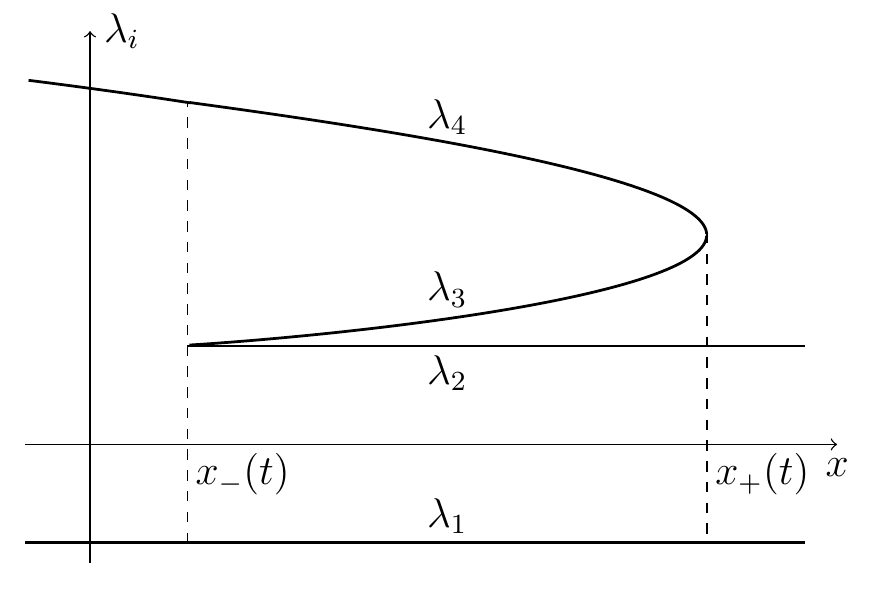}}
\caption{Riemann invariants $\la_i$ in a dispersive shock wave as
functions of coordinate $x$ at fixed instant $t$. }
\label{fig2}
\end{figure}

At instant $t$ the DSW occupies the spatial region
\begin{equation}\label{eq32}
  x_-(t)\leq x\leq x_+(t),
\end{equation}
matching to the smooth solution (\ref{eq30}) at its boundary
point. Comparison of velocity $v_+$ in expression (\ref{eq26})
with limiting expression (\ref{eq13}) shows that the DSW is
transformed into expression (\ref{eq30}) for $\la_4=\la_+$
at boundary $x=x_-(t)$; coefficients $A_k$ in this case must be
chosen so that the right-hand side of relation (\ref{eq22})
with $i=4$ be equal to the right-hand side of relation (\ref{eq30}).
Further, the solution to the Whitham equations is transformed
into a harmonic wave at the small-amplitude
edge $x=x_+(t)$, if $\la_2=\la_+=\sqrt{\rho_0}$ along the entire DSW,
and we have $\la_3=\la_4$ at point $x_+(t)$. Since $\la_-=-\sqrt{\rho_0}$
at both edges of the SDW, the condition of matching of $\la_1$ to $\la_-$
at the DSW edges can be satisfied by setting $\la_1=-\sqrt{\rho_0}$
along the DSW. Therefore, Whitham equations (\ref{eq10}) with $i=1,2$
are satisfied by constant solutions
\begin{equation}\label{eq33}
  \la_1=-\sqrt{\rho_0},\qquad \la_2=\sqrt{\rho_0},
\end{equation}
and only two Riemann invariants $\la_3$ and $\la_4$, which
satisfy the boundary conditions
\begin{equation}\label{eq34}
\begin{split}
  &\la_4=\la_+,\quad \la_3=\la_2,\quad w_4=-(\la_4-\la_2)^n,\\
   &(m=1),\quad x=x_-(t)
  \end{split}
\end{equation}
and
\begin{equation}\label{eq35}
  \la_3=\la_4,\qquad (m=0),\quad x=x_+(t)
\end{equation}
vary along the DSW. These conditions define the solution completely.
As a result, the dependence of the
Riemann invariants on coordinate $x$ for a fixed value of $t$
has the form shown in Fig.~\ref{fig2}.
It should be noted that
waves with two variable Riemann invariants were
called quasi-simple and were studied for the first time
in \cite{gkm-89} in the theory of the KdV equation. Taking
relations (\ref{eq33}) into account, we can find the first three
coefficients in Eq.~(\ref{eq20}) in the form
\begin{equation}\label{eq36}
  \begin{split}
  & W_0=1,\quad W_1=\frac12(\la_3+\la_4),\\
  & W_3=\frac18(4\la_2^2+3\la_3^2+2\la_3\la_4+3\la_4^2),\\
  & W_3=\frac1{16}(\la_3+\la_4)(4\la_2^2+5\la_3^2-2\la_3\la_4+5\la_4^2),
  \end{split}
\end{equation}
the knowledge of these coefficients is sufficient for
analyzing typical cases with $n=2$ and $n=3$.

\subsection{The case with $n=2$}

Since formulas (\ref{eq36}) are polynomial, quadratic
function $-(\la_4-\la_2)^2$ can be written in the form of a
linear combination of the first three expressions (\ref{eq36})
with coefficients
\begin{equation}\label{eq37}
  A_0=-\frac4{15}\la_2^2,\quad A_1=\frac{16}{15}\la_2,\quad A_2=-\frac{8}{15}.
\end{equation}
Then formulas (\ref{eq22}) with $i=3,4$ and with these values
of the coefficients define implicitly the dependencies
of $\la_3$ and $\la_4$ on $x$ and $t$, which solves in principle the
problem in this particular case. At the soliton edge,
these formulas are transformed into
\begin{equation}\label{eq38}
    \begin{split}
    &x-\frac12(\la_4+\la_2)t=-\frac15(\la_4-\la_2)^2,\\
    & x-\frac12(3\la_4-\la_2)t=-(\la_4-\la_2)^2,
    \end{split}
\end{equation}
which gives the relation between $t$ and $\la_4$ at this
boundary:
\begin{equation}\label{eq39}
  t=\frac45(\la_4-\la_2),\qquad \la_4=\la_2+\frac54t.
\end{equation}
Substituting the resultant value of $\la_4$ into any relation
from (\ref{eq38}) we obtain the law of motion of the soliton edge:
\begin{equation}\label{eq40}
  x_-(t)=\la_2t+\frac5{16}t^2.
\end{equation}
At the small-amplitude edge for $\la_3=\la_4$ both formulas(\ref{eq22}) with $i=3,4$
are transformed into the same relation
\begin{equation}\label{eq41}
  x_+-\left(2\la_4-\frac{\la_2^2}{\la_4}\right)t=
  -\frac8{15}(\la_4-\la_2)^2\left(3+\frac{2\la_2}{\la_4}\right).
\end{equation}
This edge moves with the group velocity corresponding to wavenumber
$k=2\pi/L$ with $L=2\sqrt{\la_4^2-\la_2^2}$, which gives for the Bogoliubov dispersion law
$\omega=k\sqrt{\la_2^2+k^2/4}$ the expression $d\omega/dk=2\la_4^2-\la_2^2/\la_4$.
Therefore, the differentiation of expression (\ref{eq41}) with
respect to $t$ with account of $dx_+/dt=d\omega/dk$
determines the dependence of $t$ on the value of $\la_4$ at this
boundary. Introducing parameter $y=\la_4/\la_2$ instead of
$\la_4$, we can write this dependence in the form
\begin{equation}\label{eq42}
  t=\frac{16\la_2}{15}\cdot\frac{(y-1)(3y^2+y+1)}{2y^2+1},\quad y\geq1,
\end{equation}
and its substitution into relation (\ref{eq41}) gives
\begin{equation}\label{eq43}
  x_+=\frac{8\la_2^2}{15}\cdot\frac{(y^2-1)(6y^2-1)}{2y^2+1}.
\end{equation}
Formulas (\ref{eq42}) and (\ref{eq43}) define parametrically the law
of motion $x_+=x_+(t)$ of the small-amplitude edge. For
$t\sim\la_0y\gg \la_2^2=\rho_0$ this law of motion asymptotically
takes the simple form
\begin{equation}\label{eq44}
  x_+(t)\approx\frac58t^2,\quad t\gg\rho_0.
\end{equation}
It should be noted that analogous expressions
obtained by solving the problem of motion of the condensate under
the action of a uniformly accelerated
piston \cite{kk-10} can be transformed to the expression
obtained above after the transfer of the breaking point
to the origin and subtracting the breaking time from $t$.

\subsection{Case with $n=3$}

\begin{figure}[th]
\centerline{\includegraphics[width=8cm]{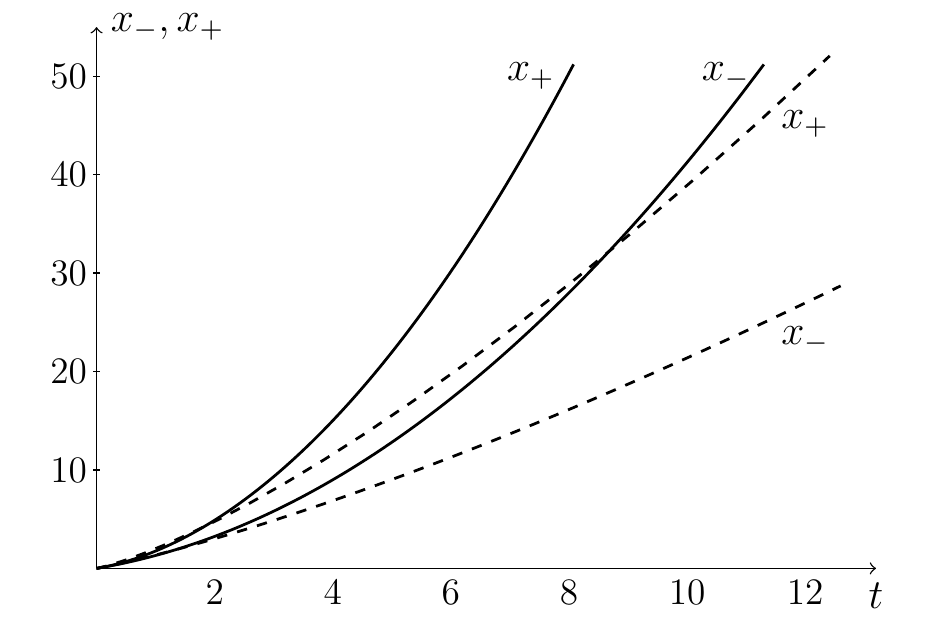}}
\caption{Motion of DSW edges for $n=2$  (solid curves) and $n=3$
(dashed curves). Background density is $\rho_0=1$. }
\label{fig3}
\end{figure}

In this case, the calculations are performed analogously.
The right-hand sides of formulas (\ref{eq22}) with
$i=3,4$ now contain function $W_3$ and condition $(\ref{eq34})$ gives
the values of the coefficients
\begin{equation}\label{eq45}
\begin{split}
  & A_0=-\frac8{35}\la_2^3,\quad A_1=-\frac87\la_2^2,\\
  & A_2=\frac{48}{35}\la_2,\quad A_3=-\frac{16}{35}.
  \end{split}
\end{equation}
At the soliton edge, solution (\ref{eq22}) is transformed into
\begin{equation}\label{eq46}
    \begin{split}
    &x-\frac12(\la_4+\la_2)t=-\frac17(\la_4-\la_2)^3,\\
    & x-\frac12(3\la_4-\la_2)t=-(\la_4-\la_2)^3,\\
    \end{split}
\end{equation}
which gives
\begin{equation}\label{eq47}
  t=\frac67(\la_4-\la_2)^2,\qquad \la_4=\la_2+\sqrt{\frac{7t}6}.
\end{equation}
The substitution of this relation into (\ref{eq46}) gives the law
of motion of the soliton edge:
\begin{equation}\label{eq48}
  x_-=\la_2t+\frac13\sqrt{\frac76}\,t^{3/2}.
\end{equation}
At the small-amplitude edge, formulas  (\ref{eq22}) with $i=3.4$ are transformed into
\begin{equation}\label{eq49}
  x_+-\left(2\la_4-\frac{\la_2^2}{\la_4}\right)t=
  -\frac{16}{35}(\la_4-\la_2)^3\left(4+\frac{3\la_2}{\la_4}\right).
\end{equation}
The matching condition for the law of motion with the
group velocity of a linear wave gives
\begin{equation}\label{eq50}
  t=\frac{48\la_2^2}{35}\cdot\frac{(y-1)^2(4y^2+2y+1)}{2y^2+1},\quad y\geq1,
\end{equation}
and the substitution into expression (\ref{eq49}) gives
\begin{equation}\label{eq51}
  x_+=\frac{16\la_2^3}{35}\cdot\frac{(y-1)^2(16y^3+14y^2-4y-5)}{2y^2+1}.
\end{equation}
For asymptotically long times $t\sim\la_0^2y^2\gg \la_2^2=\rho_0^2$ we
obtain
\begin{equation}\label{eq52}
  x_+(t)\approx \frac13\sqrt{\frac{35}6}\,t^{3/2},\quad t\gg\rho_0^2.
\end{equation}
The laws of motion of the DSW edges as functions of
time for $n=2,3$  is shown in Fig.~\ref{fig3}.
For short times, the
motion with the velocity of sound $\sqrt{\rho_0}$,
prevails, while,
for long times, a transition to asymptotic laws occurs.

\begin{figure}[ht]
\centerline{\includegraphics[width=8cm]{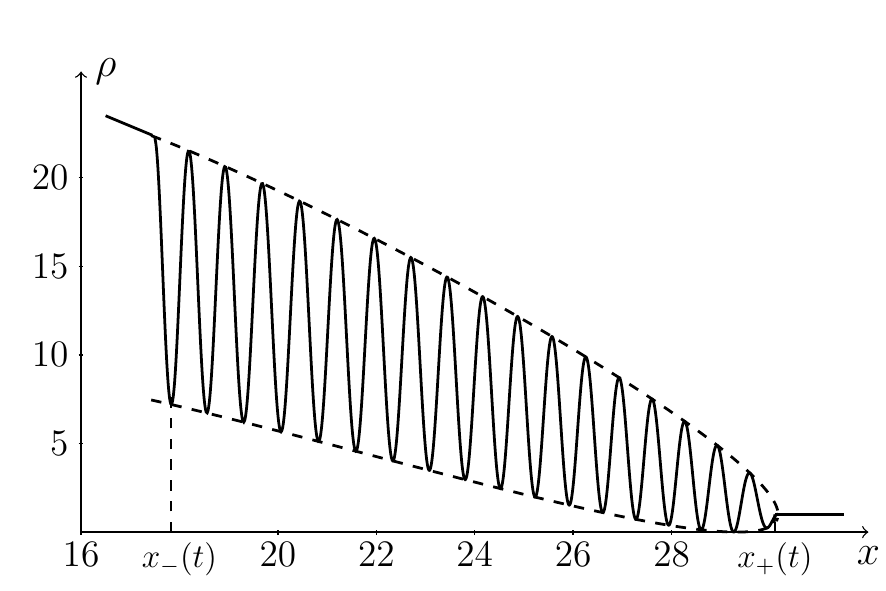}}
\caption{Condensate density profile in a dispersion shock
wave during wave breaking with $n=2$, $\rho_0=1$. The evolution time is $t=6$.}
\label{fig4}
\end{figure}

General formulas (\ref{eq22}) together with the specific
expressions for functions $w_i^{(k)}$, obtained above with the help
of formulas (\ref{eq21}), (\ref{eq36}) and coefficients (\ref{eq37}) ($n=2$)
and (\ref{eq45}) ($n=3$), make it possible to calculate
$\la_3$ and $\la_4$ as functions of $x$ and $t$, so that their substitution into expressions
(\ref{eq4}) and (\ref{eq5})  gives the density and
velocity distribution profiles in a DSW (Fig.~\ref{fig4}). The
envelopes of the condensate density in the DSW are
calculated by the formulas
$$
\rho_{max}=\frac14(\la_4-\la_3+2\sqrt{\rho_0})^2,\quad
\rho_{min}=\frac14(\la_4-\la_3-2\sqrt{\rho_0})^2.
$$

Although the formulas are complicated with
increasing exponent $n$,
important DSW characteristics
such as the laws of motion of the edges can be determined without
detailed analysis of the complete solution (at least,
in the asymptotic limit $t\gg \rho_0^n$).
In the next sections, we will consider this problem.

\section{Law of motion of the soliton edge}

Combining relations (\ref{eq34}) with the limiting expression of formula (\ref{eq21}),
we can write the boundary condition at the soliton edge in the form of a differential
equation for the function $W=W(-\la_2,\la_2,\la_2,\la_4)$,
which depends only on $\la_4$:
\begin{equation}\label{eq53}
  W+2(\la_4-\la_2)\frac{dW}{d\la_4}=-(\la_4-\la_2)^n,
\end{equation}
the solution to this equation is
\begin{equation}\label{eq54}
  W(-\la_2,\la_2,\la_2,\la_4)=-\frac1{2n+1}(\la_4-\la_2)^n,
\end{equation}
where the integration constant is chosen so that time $t$
in subsequent formulas tends to zero for $\la_4\to\la_2$. Since $w_3=W$ at this boundary,
formulas (\ref{eq22}) give
\begin{equation}\label{eq55}
    \begin{split}
    &x-\frac12(\la_4+\la_2)t=-\frac1{2n+1}(\la_4-\la_2)^n,\\
    & x-\frac12(3\la_4-\la_2)t=-(\la_4-\la_2)^n,
    \end{split}
\end{equation}
which is in conformity with relations (\ref{eq38}) ($n=2$) and (\ref{eq46}) ($n=3$). This gives
\begin{equation}\label{eq56}
  t=\frac{2n}{2n+1}(\la_4-\la_2)^{n-1},
\end{equation}
and
\begin{equation}\label{eq57}
  x_-(t)=\la_2t+\frac{n-1}{2n}\left(\frac{2n+1}{2n}\right)^{\frac1{n-1}}t^{\frac{n}{n-1}}.
\end{equation}
These expressions generalize the formulas obtained
above to arbitrary integer values of $n>1$.

It should be noted that, in fact, we can find the law
of motion of the soliton edge for an arbitrary monotonic dependence
of the initial distribution of Riemann invariant $\la_+$ of the form
$w=w(\la_+-\sqrt{\rho_0})$, by resorting to the considerations used in
\cite{gkm-89} for deriving
the law of motion of the small-amplitude edge for
wave breaking in the theory of the KdV equation.
Indeed, Whitham equations (\ref{eq10}) with $i=3,4$
in the classical hodograph method are transformed into linear differential
equations for functions $x=x(\la_3,\la_4)$ and $t=t(\la_3,\la_4)$,
one of which for  $\la_3\to\la_2$ becomes
\begin{equation}\label{eq58}
  \frac{\prt x}{\prt\la_4}-\frac12(\la_2+\la_4)\frac{\prt t}{\prt\la_4}=0.
\end{equation}
On the other hand, the solution at this boundary must
match to the smooth solution, which gives
$$
x-\frac12(3\la_4-\la_2)t=w(\la_4-\la_2).
$$
The differentiation of this relation with respect to $\la_4$
leads to one more differential equation
\begin{equation}\label{eq59}
  \frac{\prt x}{\prt\la_4}-\frac32t-\frac12(3\la_4-\la_2)\frac{\prt t}{\prt\la_4}=
  \frac{\prt w}{\prt\la_4}.
\end{equation}
Eliminating $\prt x/\prt\la_4$ from Eqs.~(\ref{eq58}) and (\ref{eq59}), we obtain
the differential equation for $t$, the integration of which
gives $(z=\la_4-\la_2)$
\begin{equation}\label{eq60}
  t=t(z)=-z^{-3/2}\int_0^z z^{1/2}\frac{dw}{dz}dz,
\end{equation}
and, hence,
\begin{equation}\label{eq61}
  x_-=x_-(z)=(\tfrac32z+\la_2)t(z)-w(z).
\end{equation}
These formulas specify the parametric dependence $x_-=x_-(t)$.

\section{Law of motion of the small-amplitude edge}

Formulas (\ref{eq41}) $(n=2)$ (\ref{eq49}) and $(n=3)$ have a simple structure
leading to the assumption that for $m\to0$ $(\la_3\to\la_4)$ and integer $n$ functions $w_3$ and $w_4$
must pass to the right-hand side of the relation
\begin{equation}\label{eq62}
  x_+-\left(2\la_4-\frac{\la_2^2}{\la_4}\right)t=
  A_n(\la_4-\la_2)^n\left(n+1+\frac{n\la_2}{\la_4}\right).
\end{equation}
Let us prove this formula in the asymptotic limit $t\to\infty$ $(\la_4\to\infty)$,
when the terms with $\la_2/\la_4$ can be
neglected. We note that in the limit $\la_2\to0$
generating function (\ref{eq19}) can be reduced to the generating function
of the Legendre polynomials (see, for example, \cite{ww})
\begin{align*}
  \begin{split}
  W&\approx\frac1{\sqrt{1-2\left(\frac{\la_3+\la_4}{2\sqrt{\la_3\la_4}}\right)
  \frac{\sqrt{\la_3\la_4}}{\la}+\left(\frac{\sqrt{\la_3\la_4}}{\la}\right)^2}}=\\
  &=\sum_nP_n\left(\frac{\la_3+\la_4}{2\sqrt{\la_3\la_4}}\right)\frac{(\la_3\la_4)^{n/2}}{\la^n},
  \end{split}
\end{align*}
i.e.,
\begin{equation}\label{eq63}
  W_n\approx (\la_3\la_4)^{n/2}P_n\left(\frac{\la_3+\la_4}{2\sqrt{\la_3\la_4}}\right).
\end{equation}
Using the recurrent formula for the derivative of the
Legendre polynomial (see \cite{ww}) we can easily prove that
\begin{equation}\label{eq64}
  \frac{\prt W_n}{\prt \la_4}=\frac{n}{\la_4-\la_3}(W_n-\la_3W_{n-1})
\end{equation}
in this approximation. To evaluate function $w_4^{(n)}$ in the
limit $m\to0$ we will prove that the following relation
holds for  $\la_3\to\la_4$:
$$
W_n-\la_3W_{n-1}\approx(\la_4-\la_3)\cdot\frac12\la_4^{n-1}.
$$
For this purpose, we note that the argument of the
Legendre polynomial in expression (\ref{eq63}), which is the
ratio of the arithmetic mean to the geometric mean,
attains its maximal value for $\la_3=\la_4$
and, hence, is
quadratic in the small difference $\la_4-\la_3=\eps$, so that $P_n(1)=1$, i.e.,
$W_n(\la_4,\la_4)=\la_4^n$. Consequently, in the
first order in $\eps$ we obtain
\begin{align*}
  \begin{split}
  & W_n(\la_4-\eps,\la_4)-(\la_4-\eps)W_{n-1}(\la_4-\eps,\la_4)\approx\\
  & \approx(\la_4-\eps)^{\frac{n}2}\la_4^{\frac{n}2}-\la_4(\la_4-
  \eps)^{\frac{n-1}2}\la_4^{\frac{n-1}2}+\eps\la_4^{n-1}\approx\\
  &\approx \frac12\eps \la_4^{n-1}.
  \end{split}
\end{align*}
In addition, we note that for $\la_2\to0$ only the highest
term with  $k=n$ is left in the sum in expression (\ref{eq22})
since coefficients $A_k$ for $k<n$  contain powers of $\la_2$ as
factors. Therefore, with account of relation.
$2(V-v_4)\approx-2\la_4$ we obtain
\begin{equation}\label{eq65}
  w_4\approx A_n(n+1)\la_4^n.
\end{equation}
For determining $w_4$ completely for $m=0$ it remains
to find $A_n$, which can easily be done using the
expansion of the higher term in the Legendre polynomial into a series
for $\la_3\to\la_2=0$ (see \cite{ww}):
\begin{equation}\label{eq66}
  \begin{split}
  W_n&\approx(\la_2\la_4)^{\frac{n}2}\cdot\frac{(2n)!}{2^n(n!)^2}
  \left(\frac{\la_2+\la_4}{2\sqrt{\la_2\la_4}}\right)^n\approx\\
  &\approx \frac{(2n)!}{4^n(n!)^2}\,\la_4^n.
  \end{split}
\end{equation}
Then the condition of matching to the smooth solution gives
\begin{align*}
w_4\approx A_n\cdot  \frac{(2n+1)!}{4^n(n!)^2}\,\la_4^n =-\la_4^n,
\end{align*}
i.e.
\begin{equation}\label{eq67}
  A_n=-\frac{4^n(n!)^2}{(2n+1)!}.
\end{equation}
Differentiating the expression
\begin{equation}\label{eq68}
  x-2\la_4t=w_4=A_n(n+1)\la_4^n
\end{equation}
with respect to $\la_4$ provided that $\prt x/\prt\la_4=0$  for a fixed $t$,
which is equivalent to the matching condition with
the group velocity at the small-amplitude edge, we
obtain
\begin{equation}\label{eq69}
  t=-\frac12n(n+1)A_n\la_4^{n-1}.
\end{equation}
Substituting $\la_4$ obtained from this expression into (\ref{eq68})
we obtain the law of motion of the small-amplitude edge in the asymptotic regime:
\begin{equation}\label{eq70}
  x_+=\frac{2(n-1)}n\left(\frac{(2n+1)!}{2^{2n-1}n(n+1)(n!)^2}\right)^{\frac1{n-1}}t^{\frac{n}{n-1}}.
\end{equation}
This formula naturally reproduces the above asymptotic laws
(\ref{eq44}) for $n=2$ and (\ref{eq52}) for $n=3$.

Expression (\ref{eq68}) confirms the validity of formula (\ref{eq62})  in the limit
$\la_2\to0$. Assuming that this formula
also holds for a finite $\la_2$,
we obtain the law of motion
of the small-amplitude boundary in parametric form:
\begin{equation}\label{eq71}
  \begin{split}
  t&=\frac{nA_n\la_2^{n-1}(y-1)^{n-1}}{2y^2+1}[(n+1)y^2+(n-1)y+1],\\
  x&=\frac{A_n\la_2^n(y-1)^{n-1}}{2y^2+1}[2(n^2-1)y^3+\\
  &+2(n^2-n+1)y^2-(n-1)^2y-n^2+n+1].
  \end{split}
\end{equation}

It should be noted that in contrast to the theory of
the KdV equations, the self-similar regime of motion
of the boundaries is realized only asymptotically for
long times
$t\gg \rho_0^{n-1}$, when the velocity of motion is
much higher than the velocity of sound in the background distribution.
However, the limiting transition to $\rho_0\to0$ is impossible in the expressions describing
the wave profile since the magnitude of $m$ in elliptic
functions vanishes for $\la_2=-\la_1\to0$.

\section{Conclusion}

Thus, the approach developed by Gurevich and
Pitaevskii makes it possible to analyze in detail the
process of DSW formation during wave breaking in the
Bose-Einstein condensate, the dynamics of which
obeys the Gross-Pitaevskii equation. The developed
theory is applicable to the initial stage of the process,
in which the smooth part of the profile can be treated
as a monotonic function of the coordinate. It should
be noted, however, that the theory of quasi-simple
waves must also describe the asymptotic stage of the
evolution of a finite-duration pulse since, analogously
to the simple wave theory, the initial pulse splits with
time into two pulses, in each of which two of four Riemann 
invariants again remain constant. Therefore, the
Gurevich-Pitaevskii approach supplemented with the
generalized hodograph method and modern method
for deriving the Whitham modulation equations
remains a powerful tool for investigating dispersion
shock waves, which are of considerable interest for
modern nonlinear physics.

\begin{acknowledgments}
  I thank M. Isoard, S. K. Ivanov and N. Pavloff for useful
  discussions.
\end{acknowledgments}


\begin{references}

\bibitem{LL-6} L. D. Landau, E. M. Lifshitz, {\it Fluid Mechanics,} Pergamon,
 Oxford (1987).

\bibitem{CF-50} R. Courant,  K. O. Friedrichs, {\it Supersonic Flow and Shock Waves},
Interscience Publishers,  New York (1948).

\bibitem{bl-54} T.~B.~Benjamin, M.~J.~Lighthill, Proc. Roy. Soc. London, A {\bf 224}, 448 (1954).

\bibitem{sagdeev} R. Z. Sagdeev, {\it Cooperative phenomena and shock waves in collisionless
plasmas}, Rev. Plasma Phys. \textbf{4},  23 (1966).

\bibitem{gp-73} A. V. Gurevich and L. P. Pitaevskii,
  Zh. Eksp. Teor. Fiz. {\bf 65}, 590 (1973) [Sov. Phys. JETP {\bf 38},
  291 (1974)].

\bibitem{whitham} G.~B.~Whitham, Proc. Roy. Soc. London, A {\bf 283}, 238 (1965).

\bibitem{potemin} G. V. Potemin, Russian Math. Surveys, {\bf 43}, 252 (1988).

\bibitem{kamch} A.~M.~Kamchatnov, {\it Nonlinear Periodic Waves and Their Modulations.
An Introductory Course}, World Scientific, Singapore (2000).

\bibitem{eh-16} G.~A.~El, M.~A.~Hoefer, Physica D {\bf 333}, 11 (2016).

\bibitem{gross} E.~P.~Gross.
{ Nuovo Cimento,} {\bf 20,} 454 (1961).

\bibitem{pit} L. P. Pitaevskii,  
Zh. Eksp. Teor. Fiz. {\bf  40,} 646 (1961) [Sov. Phys. JETP {\bf 3,} 451.

\bibitem{tsuzuki-1971} T.~Tsuzuki,
J. Low Temp. Phys., {\bf 4,} 441 (1971).

\bibitem{gk-87}  A.V. Gurevich and A.L. Krylov, 
Zh. Eksp. Teor. Fiz. {\bf 92,} 1684 (1987) [Sov. Phys. JETP,  {\bf 65,}  944 (1987)].

\bibitem{zs-73} V. E. Zakharov and A. B. Shabat, Zh. Eksp. Teor. Fiz., {\bf 64}, 1627 (1973)
[Sov. Phys. JETP, {\bf 37,} 823 (1973).

\bibitem{fl-86}  M.~G.~Forest and J.~E.~Lee,
in {\it Oscillation Theory,
Computation, and Methods of Compensated Compactness,} Eds. C. Dafermos et al,
IMA Volumes on Mathematics and its Applications {\bf 2,} (Springer, N.Y., 1986).

\bibitem{pavlov-87} M. V. Pavlov, 
Teor. Mat. Fiz. {\bf 71,} 351 (1987) [Theor. Math. Phys., {\bf 71,} 584 (1987)].

\bibitem{eggk-95}  G.~A.~El, V.~V.~Geogjaev, A.~V.~Gurevich, A.~L.~Krylov,
Physica D, {\bf 87,} 186 (1995).

\bibitem{kku-02} A.~M.~Kamchatnov, R.~A.~Kraenkel, B.~A.~Umarov,
Phys. Rev. E,  {\bf 66},  036609 (2002).

\bibitem{cornell} E.~A.~Cornell, Talk at NATO Advanced Workshop ``Nonlinear
Waves: Classical and Quantum Aspects,'' Lisbon, 2003.

\bibitem{hoefer} M.~A.~Hoefer, M.~J.~Ablowitz, I.~Coddington,
E.~A.~Cornell, P.~Engels, V.~Schweikhard,
{ Phys. Rev. A}   {\bf 74,} 023623 (2006).

\bibitem{kgk-04}  A.~M.~Kamchatnov,  A.~Gammal, R.~A.~Kraenkel,
Phys. Rev. A  {\bf 69,} 063605 (2004).

\bibitem{hae-08} M.~A.~Hoefer, M.~J.~Ablowitz, P.~Engels,
Phys. Rev. Lett. {\bf 100}, 084504 (2008).

\bibitem{kk-10} A.~M.~Kamchatnov, S.~V.~Korneev,
Zh. Eksp. Teor. Fiz., {\bf 137,} 191 (2010) [JETP, {\bf 110,} 170 (2010)].

\bibitem{tsarev} S. P. Tsarev, Math. USSR Izv. {\bf 37}, 397 (1991).

\bibitem{gke-92}  A. V. Gurevich, A. L. Krylov and G. A. El,
  Zh. Eksp. Teor. Fiz. {\bf 101}, 1797 (1992) [Sov. Phys. JETP
  \textbf{74}, 957 (1992)].

\bibitem{wright} O.~Wright, Commun. Pure Appl. Math., {\bf 46}, 421 (1993).

\bibitem{tian} F.~R.~Tian, Commun. Pure Appl. Math., {\bf 46}, 1093 (1993).

\bibitem{gkm-89} A. V. Gurevich, A. L. Krylov, and N. G. Mazur,
  Zh. Eksp. Teor. Fiz. {\bf 95}, 1674 (1989) [Sov. Phys. JETP {\bf
    68}, 966 (1989)].

\bibitem{ww} E. T. Whittaker and G. N. Watson, {\it A Course of Modern Analysis},
CUP, Cambridge (1927).

\end{references}
\end{document}